%% file: main.tex
\documentclass[10pt,twocolumn,letterpaper]{article}

\usepackage{cvpr}
\usepackage{times}
\usepackage{epsfig}
\usepackage{graphicx}
\usepackage{float}
\usepackage{amsmath}
\usepackage{amssymb}

\usepackage[breaklinks=true,bookmarks=false]{hyperref}

\cvprfinalcopy 

\begin{document}

\title{Economic Impact of Discoverability of Localities and Addresses in India}

\author{Dr. Santanu Bhattacharya\\
\and
Sai Sri Sathya\\
\and
Dr. Kabir Rustogi\\
\and
Dr. Ramesh Raskar
}

\maketitle

\begin{abstract}

    Most of the earth's population has a poorly defined addressing system, thus having a poorly discoverable residence, property or business locations on a map. Easily discoverable addresses are important for improving their livelihood, business-incomes, and even service delivery. The economic cycle based on discoverable addresses is self-reinforcing: consumers independently identify and adopt such addresses according to their own convenience and businesses use algorithms or third-party services to resolve these addresses into geocodes to help better identify their customers' locations. \par

    Our paper analyses from the top two industries in India: Logistics and Financial Services, indicate that the lack of a good addressing system costs India \$10-14B annually. As the Indian economy is expected to grow rapidly, the businesses would proportionately grow, causing the total costs to grow further. We, therefore, need to consider a dramatically new approach to modernize the addressing systems to bring in efficiency.
 
\end{abstract}

\section{Introduction}
\input{intro}

\section{Economic Impact on Industry}
\input{case1}
\input{case2}

\section{Pan-India Economic Impact}
\input{panindia}

\section{Conclusion}
\input{conclusion}

\section{About the Authors}
\input{about}

{\small
\bibliographystyle{ieee}
\bibliography{main}
}

\end{document}

%% file: intro.tex
Out of seven billion inhabitants on Earth, approximately 75\% \cite{1} do not have proper addresses that allow their houses, properties or businesses to be located on a map with reasonable precision. To illustrate, consider the two addresses of one of our authors in two different continents: \par
\begin{figure}[h]
\begin{center}
   \includegraphics[width=0.9\linewidth]{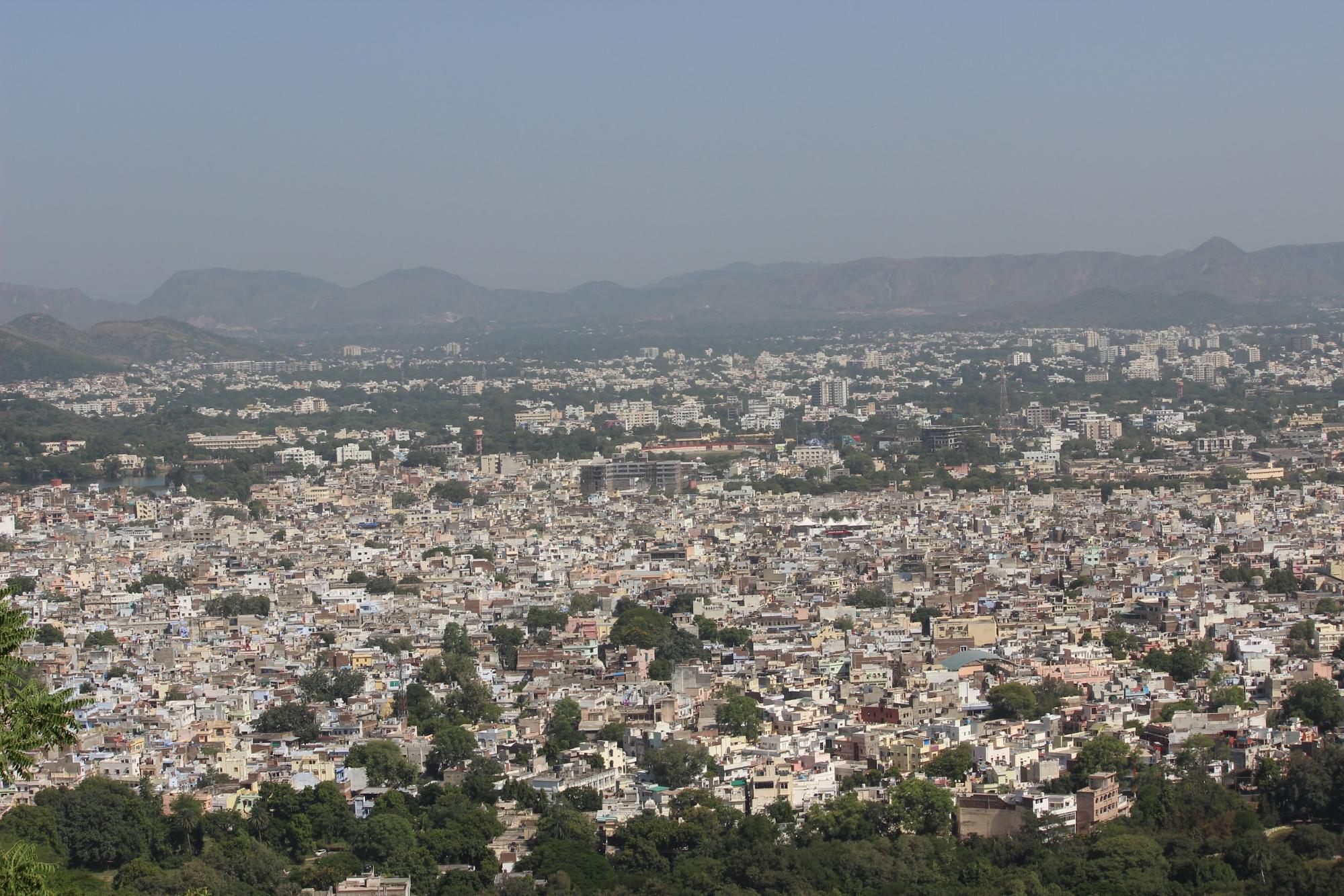}
\end{center}
   \caption{Every place deserves an address. Seen here is the city of Jaipur in India}
\label{fig:intro1}
\end{figure}

\begin{itemize}
    \item {[Name]}, 7116 Via Correto Dr, Austin, TX 78749: the location for this address is readily available and any navigation system can take you to the doorstep in day or night; or in good or bad weather! This is an example of a structured address. 
    \item {[Name]}, College Tilla, PO Agartala College, Agartala 799004: Google Maps resolves this address to an area of roughly 76 sq. km. in the city of Agartala. This address is unstructured. 
    \item Adding a landmark to the above address, ``College Tilla \textbf{near College Tilla Lake}" narrows the answer to an area of approximately 3 sq. km. Upon reaching the place, and depending on how much detailed information about the occupant or the house or the location is known (e.g. whose son/ brother the occupant is, what's his age, color of his house and if is located by the old mango tree etc.), one would take a additional 15-45 minutes, provided it's not late night or raining. Moreover, landmark-based addressing is infrequent, incomplete and also inconsistent. 
\end{itemize}

\begin{figure}[h]
\begin{center}
   \includegraphics[width=0.9\linewidth]{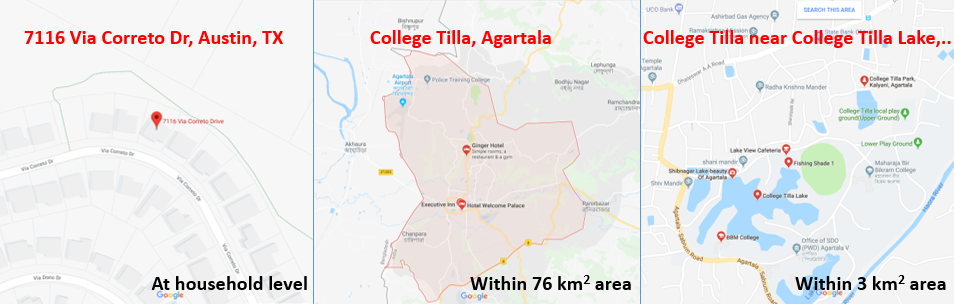}
\end{center}
   \caption{Most addresses in the developed world resolves within a narrow area, where those in India can resolve anywhere from town to a locality to (rarely) at a house level}
\label{fig:intro2}
\end{figure}

These are not merely one-off experiences that cause inconvenience. The inability to provide an accurate location for each and every address, impacts the livelihood of residents in many ways. It inhibits growth of their local trades such as salons, bakeries or food stalls. It reduces availability of amenities such as creation of banks accounts and delivery of goods and services (e.g., e-commerce) and delays emergency services such as fire brigades and ambulances.

%% file: case1.tex
\subsection{Logistics and Transportation}

The inability to locate an address within a reasonable accuracy, as demonstrated in the previous example, hampers a transporter's ability to deliver shipments on time and without incurring additional costs. Consider the e-commerce industry, or any industry where goods are delivered to an address. In absence of proper geocodes, most companies in India, ``sort" the packages or goods by pincode, since pincode is the only numeric location depictor that appears in most written addresses in India.\par

While this sounds like a logical solution, e.g., to have deliveries sorted by pincodes and even creating facilities (that store the shipments for delivery or return) according to the demand volumes those exist in respective pincodes. However, two major, practical problems arise while implementing this idea:

\begin{itemize}
    \item About 30-40\% \cite{2} of the pincodes in India are written incorrectly, leading to shipments being misrouted and requiring manual intervention for eventual routing to the correct pincode 
    \item The average area a pincode covers 179 sq. km. with about 135,000 households and over 100,000 business, educational institutions, government buildings etc. Sorting deliverables for more than a quarter million addresses, where only 30\% of those addresses are structured, poses many challenges 
\end{itemize}

We will pick an area within the Logistics and Transportation industry to illustrate the challenge in detail. Consider an e-commerce, which is growing at a 30 CAGR \cite{3}, driven by rising income, consumption and digitization \cite{4}, and it is expected to continue to do so for the foreseeable future. Indian consumers expect products to be delivered at their doorsteps for free, which causes a unique burden on the e-commerce companies not seen in most parts of the world. \par

When a product from an e-commerce site is ordered online, the merchandize is picked up from a seller or a warehouse and brought to a processing, where it is sorted for the destination city- a process known as the ``first mile operation". It is then transported between origin and destination cities in a ``line haul" that involves long-distance transportation such as truck, air etc. In the ``last mile operation", the merchandize goes to the delivery center from where it is delivered to the shopper's house. Figure 3 illustrates this process.

\begin{figure}[h!]
\begin{center}
   \includegraphics[width=0.9\linewidth]{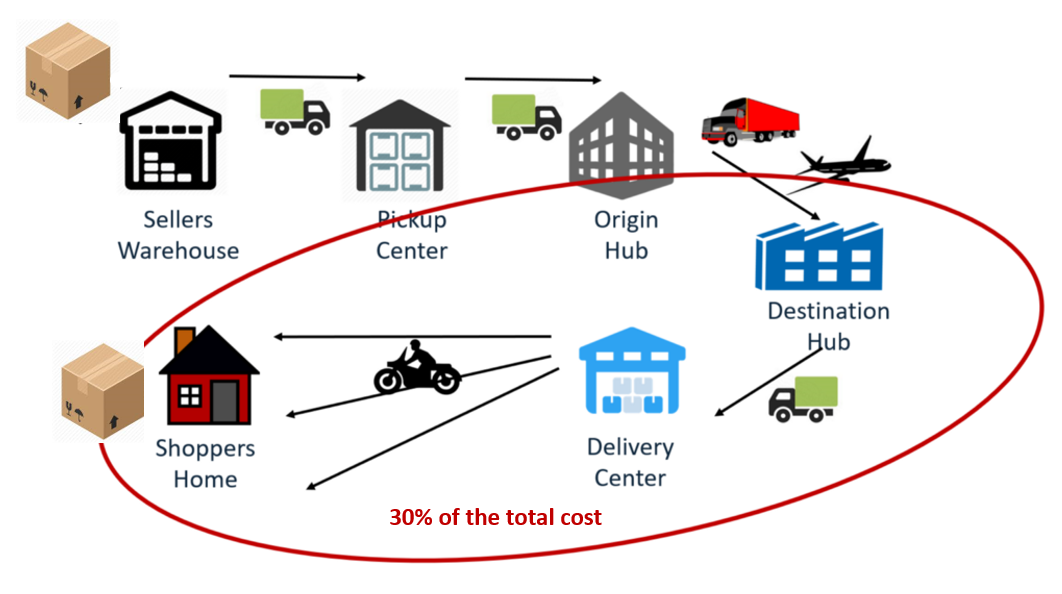}
\end{center}
   \caption{The ``last mile cost" in India is ~30\% of the total cost of delivery}
\label{fig:case11}
\end{figure}

In western countries, structured addresses lead to a relatively accurate geocoding and consequently the last mile cost is about 10-12\% of the total cost \cite{5}. In India, the same cost is ~35\% of the total cost of delivery; notwithstanding India's low cost of labor. The extra cost comes from the longer time that a driver takes in locating an address- to drive to an address, stopping multiple times to either call the recipient or ask someone on the road or nearby shops for location and directions, time wasted on additional kilometers driven in search of the address etc. \par

Figure 4. Illustrates the challenges of delivering to addresses that cannot be disambiguated at the house level. In absence of a lat-long for the desired address, addresses are sorted based on pincodes and all the packages for one pincode are sorted, stored and delivered from one (or, sometimes, two or more) delivery center(s). Typical pincode-based sorting centers, located at the orange pin location can have a delivery ``throw" (radius) of 4-20 km. In absence of understanding load distribution based on addresses, such centers' locations often tend to be imbalanced. For example, in this case, while one delivery biker drives about 22 km, the other drives about 104 km, almost 5 times the distance covered by the first one. Such problems significantly hamper the initiatives to improve productivity and reduce costs at these centers. \par

\begin{figure}[h]
\begin{center}
   \includegraphics[width=0.9\linewidth]{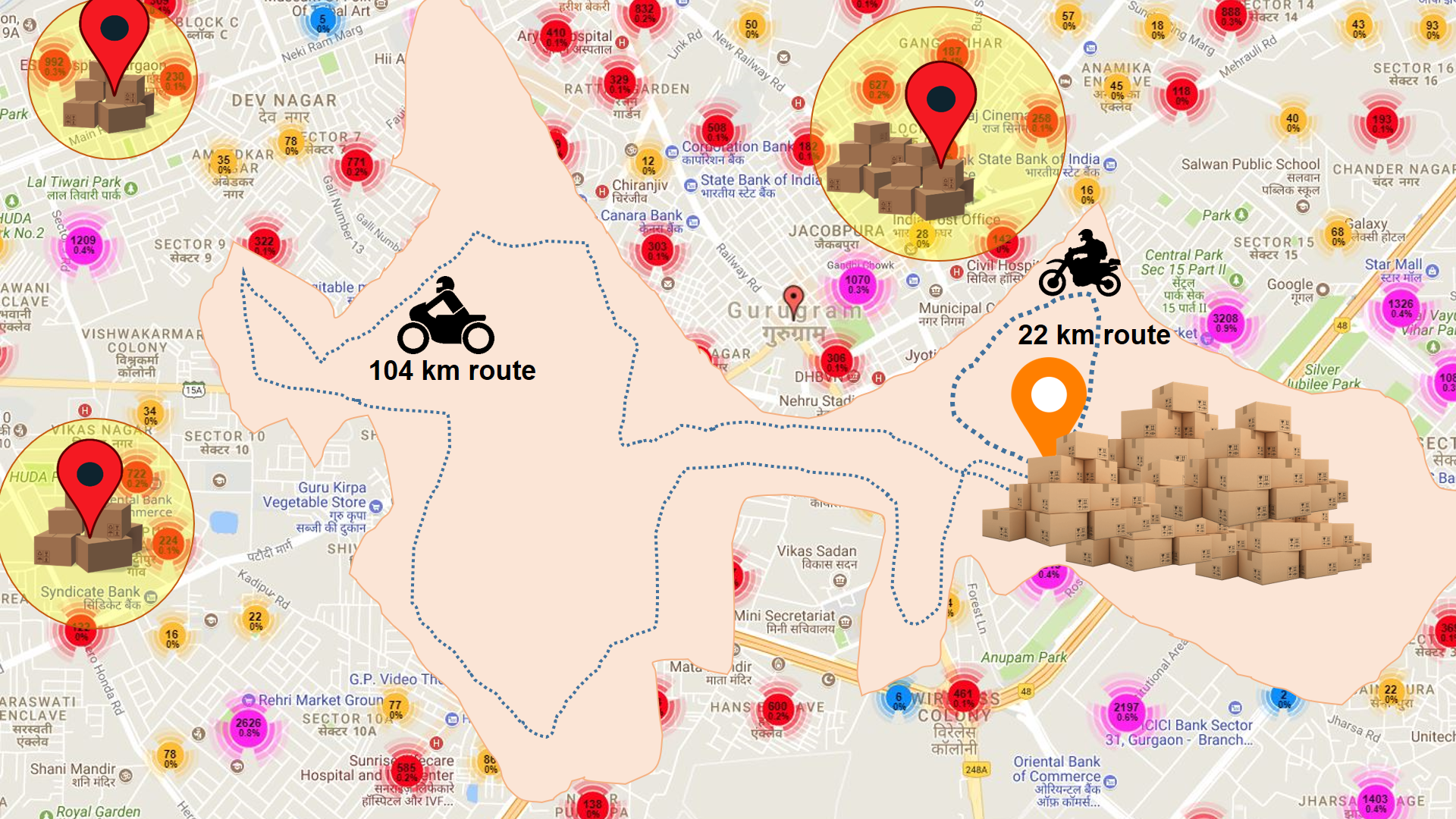}
\end{center}
   \caption{The difference between a pincode-based sorting and local-address-based sorting. In this picture, the light orange area shows a typical pincode boundary. In this delivery center productivity can vary widely as optimal route planning becomes complex, as depicted by two delivery bikers' routes}
\label{fig:case12}
\end{figure}

If the addresses, on the other hand, can be disambiguated down to a household level, then each individual locality can have a small delivery center, much like it is common to have pharmacy, grocery store or a restaurants in most areas. In such cases, the ``throw" of the delivery center goes down to an average of 1-3 km. The bikers typically cover small distances in a shift and multiple shifts can be run in a day to match the schedule of multiple trucks that can bring loads throughout the day from the destination hubs. In Figure 4, the locations for such small delivery centers are represented by red pins and areas they cover are shown in yellow. This results in a significant improvement in productivity at the smaller centers that can do address-based sorting. \par

Moreover, more granularity in the address geocodes, also allows us to perform route optimisation and provide system driven routes for the delivery boys. We provide the case of Delhivery, one of India's leading logistics providers for e-commerce companies. \par

At Delhivery, a switch from a pincode based to a locality based sorting has improved the productivity of the last mile operation by 40-60\%, depending on the type/complexity of the addresses and size/shape of the locality. \par

\begin{table}[h]
\begin{center}
 \includegraphics[width=0.9\linewidth]{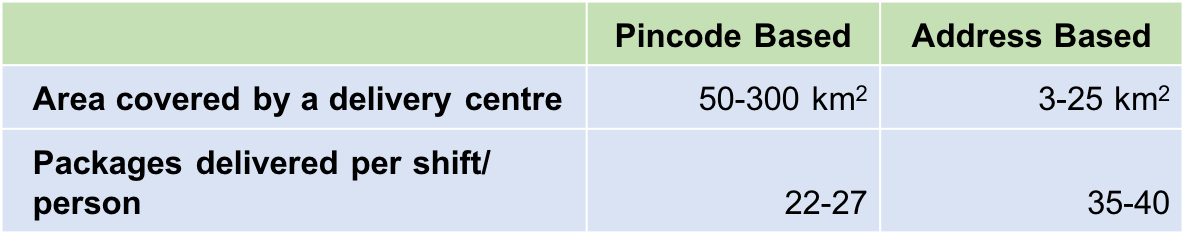}
\end{center}
   \caption{An address-based sorting can result in a 40-60\% better productivity}
\label{fig:case13}
\end{table}

This high last-mile cost disproportionately affects a company's bottom-line. In a simplistic analysis in Table 2, we demonstrate that a better geocoding which reduces the last-mile cost by 40\%, well within the reach of current technology, can improve the profitability of an e-commerce company. \par

\begin{table}[h]
\begin{center}
\includegraphics[width=0.9\linewidth]{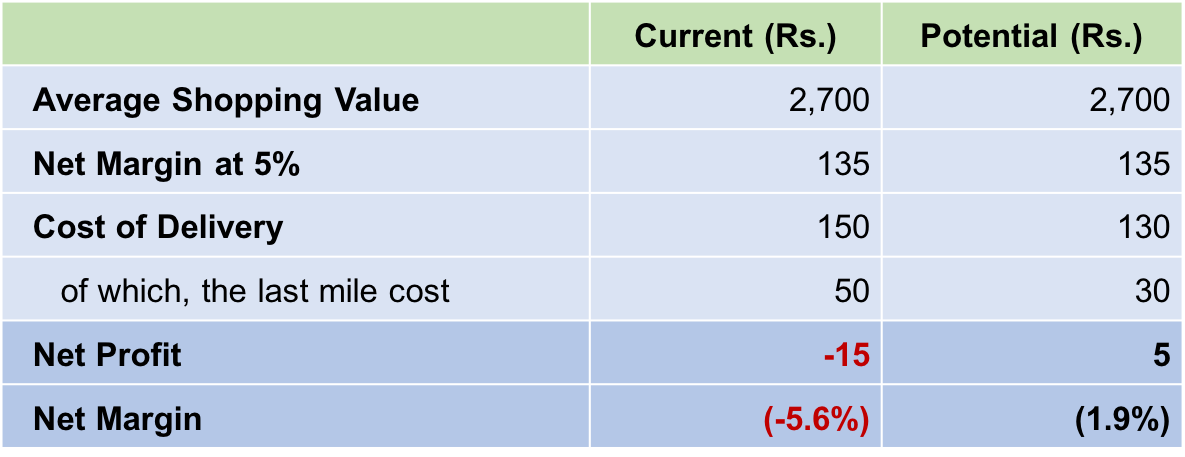}
\end{center}
   \caption{Illustration: An improvement in the last mile cost can swing the profitability of an e-commerce business}
\label{fig:case14}
\end{table}

Even for a small industry in India, such as e-commerce delivery, which is estimated to be a 5,000 Crore (~\$775M) business annually (as of 2017), the annual cost savings from a better addressing scheme is about 650 crore (~\$100M). \par

For the Logistics and transportation industry, the same framework can be used for different types of goods and services transportations. When a shipment is moved from one city to another, the line-haul segment, which is typically the inter-city transport, is not much impacted by the lack of proper geocoded addresses. However, both the first mile (pickup from a client or a distribution center, for example) or the last mile (delivery to a house or business) are impacted significantly by the inability to resolve an address, whether a bike, truck or a bicycle is used for performing that operation. \par

%% file: case2.tex
\subsection{Loan and Financial Services}

India is a credit-deprived country where 642 million people, a staggering 53\%, are excluded from formal financial products such as loans, insurance and other forms of credits and financial services. Even among those who are engaged in trades or small businesses, 48\% cannot access formal credits or loans. The impact on the economy is significant. McKinsey estimates that the payoff for digital financial services in India by 2025 can be \$700 billion and it can create an additional 21 million jobs \cite{6}. \par

The reasons for the paucity of credit are many: lack of verifiable identity (akin to social security number in the USA), absence of proof of formal income in a largely cash-driven economy and complexity of disambiguating one's location, be it home, or place of business. \par

This has started to change in the past few years. Government's initiative to provide biometric identity to all Indians (``Aadhaar"), has for the first time in history, given over 95\% Indians a verifiable identity. Additionally, initiatives to open over 100 million bank accounts for the financially disadvantaged and push for digital transactions have pursued many startups to consider providing loan and credit services to the formerly unserved population. \par

Consequently, in the past two years, funded by large venture capital investments in financial technology (``fintech") companies, over 100 startups have started providing services for connecting borrowers and lenders. \par 

Figure 5 shows a typical process promised by one of such services. The process is reasonably straightforward. Once a user applies online or through the app and selects a product, they are typically asked for 5-8 sets of documents:

\begin{itemize}
    \item A proof of \textbf{identity} such as Aadhar, passport or a voter ID card 
    \item A proof of \textbf{address} such as a lease or house ownership documents, i.e., sales deed
    \item Proof of \textbf{income} such as paycheck, tax returns or business earning 
    \item Proof of \textbf{educational qualifications}, particularly for students
    \item Proof of \textbf{age} for loan eligibility
    \item \textbf{Employment} verification, such as a certified letter from the employer
    \item \textbf{Bank} statements
    \item Additional documentation around \textbf{proof of residence} (utility bills such as phone, water or electricity) or a letter from the employer in the official letterhead, especially if original identity documents such as Aadhaar card or passport have been issued in another state
\end{itemize}
    
\begin{figure}[h]
\begin{center}
   \includegraphics[width=0.9\linewidth]{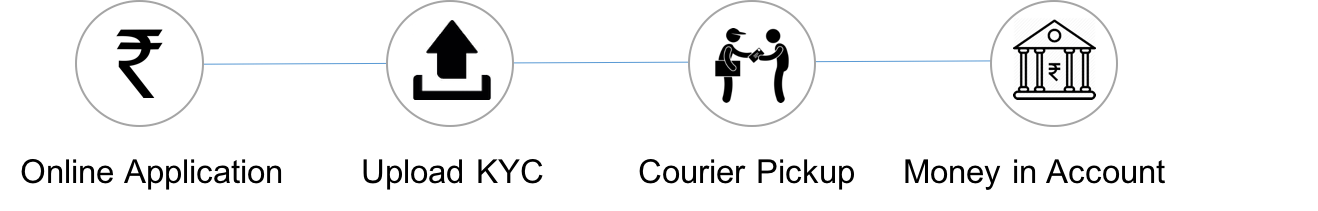}
\end{center}
   \caption{Typical loan generation process promised by one of India's many digital loan or financing start-ups}
\label{fig:case21}
\end{figure}
    
A courier picks these documents from the borrower. These documents are then scanned and saved in a database, from which it is compared against the information provided by the borrower on the loan application. Since the documents are typically paper-documents, an Optical Character Recognition (OCR) system is used for directly transcoding the information to a database. \par

This process works well for about 60-70\% of borrowers, especially in large cities. From our survey with leading firms, we estimate that about ~70\% of the documents are considered a ``match" and go to the next step for loan processing, e.g, loan eligibility analysis, approval of loans etc., albeit with only a certain percentage of applicants being eligible for loan. \par

However for the ~30\% of the applications, the address provided by the applicant in the application does not match the documents. To understand why, consider the following addresses for the same house, in Figure 6. \par

\begin{figure}[h]
\begin{center}
   \includegraphics[width=0.9\linewidth]{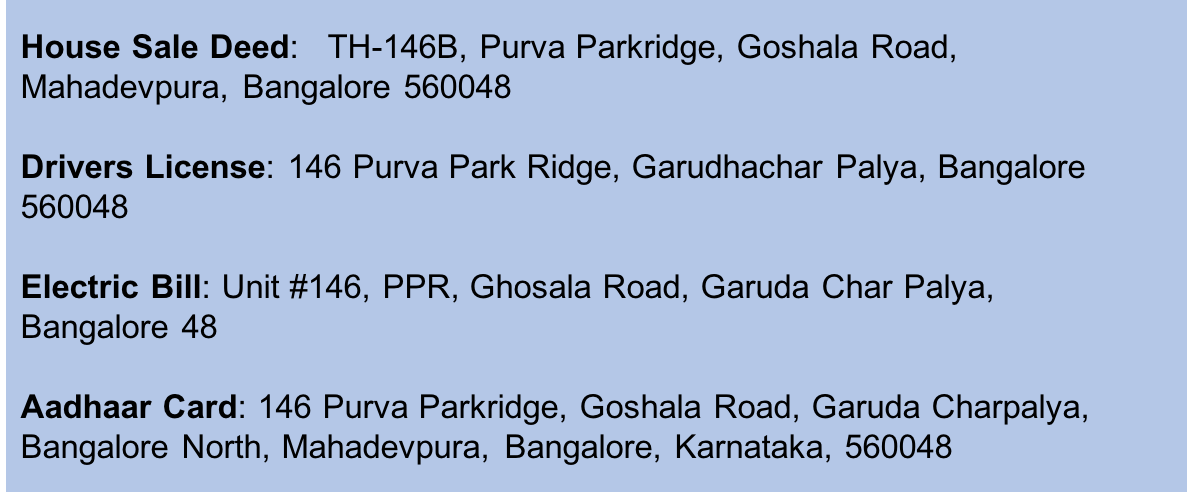}
\end{center}
   \caption{Same address written in multiple formats in different documents}
\label{fig:case22}
\end{figure}

Consider the different variations:
\begin{itemize}
    \item The house number has been written in \textbf{three} formats. ``TH-146B", ``146", as well as ``Unit 146" 
    \item The community has been described as both ``Purva Parkridge" and ``Purva Park Ridge", and has been abbreviated as ``PPR" again, in \textbf{three} different ways
    \item The road name has been spelled in two ways as ``Goshala Road" and ``Ghosala Road" and omitted in one all-together 
    \item The locality is described as ``Garuda Char Palya" and ``Garuda Charpalya". In one document, a neighboring community, ``Mahadevpura" has been substituted
\end{itemize}

The one in Aadhar card seems to have taken the path of safest approach, adding both the localities of ``Mahadevpura" and ``Garuda Charpalya" to the same address. In other words, just four different sources of official address verification documents can produce over 50 combinations. \par

It is therefore not surprising that the addresses provided in documents often do not match. Since the processing of these information happen in a centralized facility, people there would have no idea about ``Purva Park Ridge" and ``PPR" being same community or ``Garuda Char Palya", with its different ways of spelling is often interchanged with its neighbouring community ``Mahadevpura". \par

For the 30\% of documents that do not match, the following process kicks in, as depicted in Figure 7. \par

\begin{figure}[h]
\begin{center}
   \includegraphics[width=0.9\linewidth]{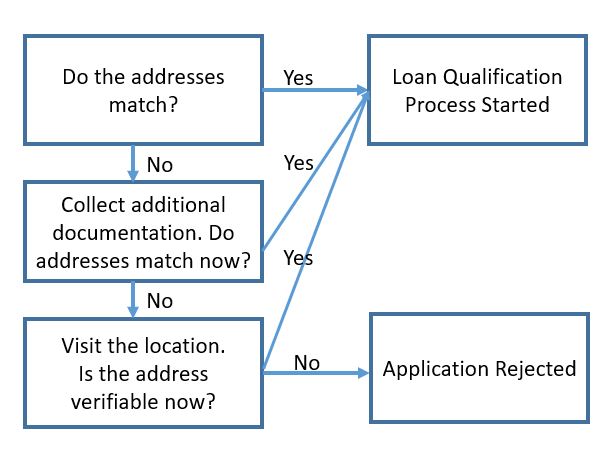}
\end{center}
   \caption{30\% of the applicants whose addresses do not match directly are either asked for additional documentations or have their addresses manually verified, adding to the time and cost for the service providers}
\label{fig:case23}
\end{figure}

This results in delayed approval of the loans by up to 5 days in best cases and even weeks or months sometimes. This affects both the borrower, who could be in urgent need of money; and the lender, who has to bear the loss of interest he could have earned until the loan is finally processed and the cost of additional verifications. \par

\begin{table}[h]
\begin{center}
\includegraphics[width=0.9\linewidth]{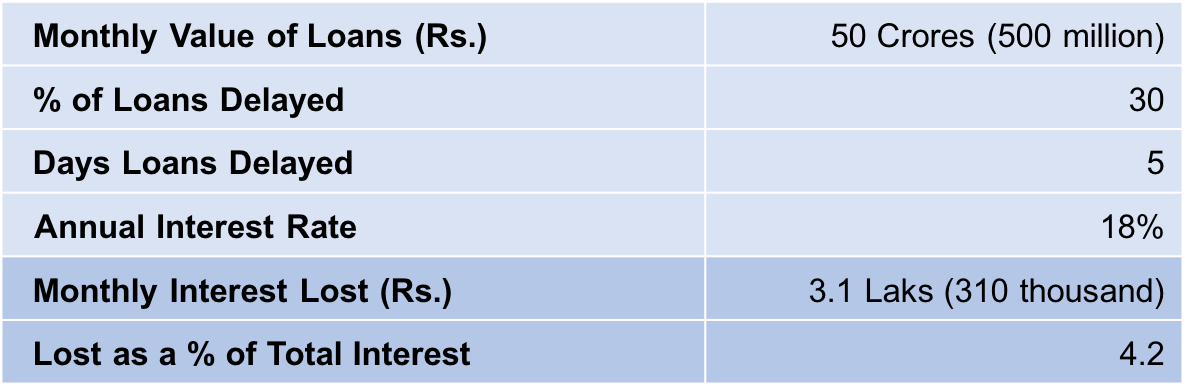}
\end{center}
   \caption{Bad addresses delay verification and approval resulting in the loss of interest to moneylenders}
\label{fig:case24}
\end{table}

Also, the place of dwelling or the  business being a key factor in the risk-assessment process of a loan, the inability to disambiguate it shows up in the risk models, raising the rate and hence, the overall cost of the loan. \par

%% file: panindia.tex
We conducted similar analysis for the top three industries- Logistics, Manufacturing (including consumer goods) and Emergency Services to derive a cost-estimate for India. Using this approach, our estimate indicate that poor addresses \textbf{cost India \$10-14B annually, ~0.5\% of the GDP}; see Table 4.

\begin{table}[h]
\begin{center}
\includegraphics[width=0.9\linewidth]{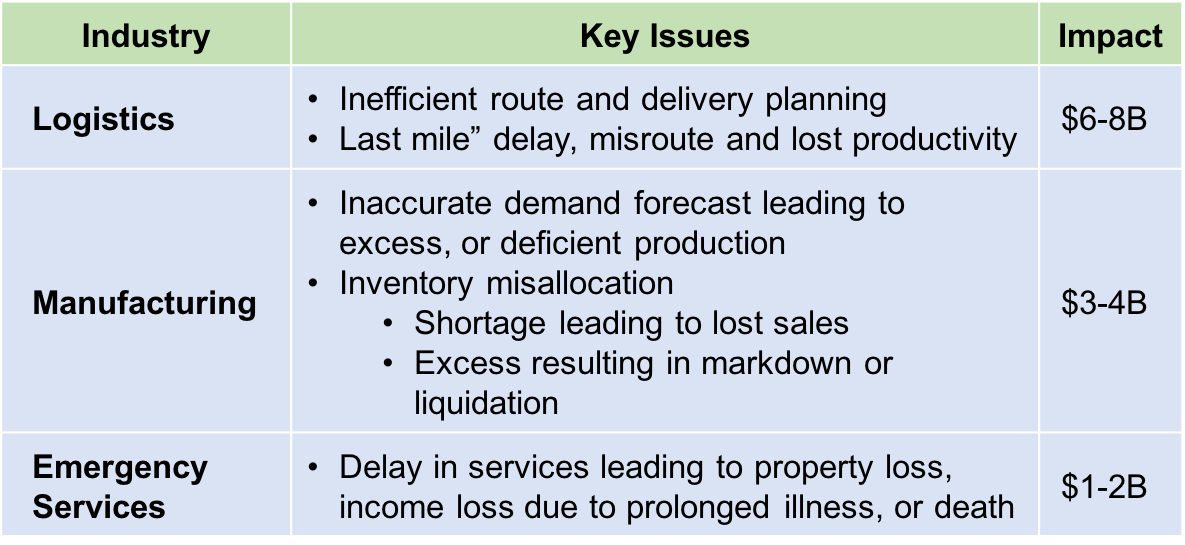}
\end{center}
   \caption{The economic cost of bad addresses in India}
\label{fig:pan1}
\end{table}

Further note that the numbers presented in Table 4 capture the cost of bad addresses, but do not include additional benefits of having better addresses like rising productivity and income gains, which lead to further growth of businesses and GDP etc.

%% file: conclusion.tex
Easily discoverable addresses are important for rapidly growing economies like India. Rather than just being a convenience, addresses are vital for driving a self-reinforcing economic cycles and therefore, improving livelihood and incomes for the next billion Indians. The consumers independently identify and adopt addresses for their own convenience while the businesses use technology or third-party services to resolve these addresses into geocodes to deliver products and services at reduced costs. However, the current addressing system in India does not lend itself to disambiguation to a reasonably accurate lat-long for most addresses.

Our case-study analyses indicate that the lack of a good addressing system costs India at least \$10-14B a year, or about 0.5\% of its annual Gross Domestic Product. As the Indian economy continues to grow in both economic output as well as variety of  new businesses and services, the costs due to lack of a proper addressing system will increase significantly. India therefore needs to consider a dramatically new approach to modernize the addressing system to bring in efficiency.

%% file: about.tex
\textbf{Dr. Santanu Bhattacharya} is scientist collaborating with Camera Culture Group at MIT Media Lab. A serial entrepreneur who has led Emerging Market Phones at Facebook, he is a former physicist from NASA Goddard Space Flight Center.

\textbf{Sai Sri Sathya} is a researcher collaborating with REDX and Camera Culture Group at MIT Media Lab and formerly at the Connectivity Lab at Facebook, focused on Emerging World Innovations.

\textbf{Dr. Kabir Rustogi} leads the Data Science team at Delhivery, India's largest E-commerce logistics company. A published \href{http://www.springer.com/gp/book/9783319395722}{author}, he was previously a Senior Lecturer of Operations Research at The University of Greenwich, UK.

\textbf{Dr. Ramesh Raskar} is Associate Professor at MIT Media Lab and leads the Emerging Worlds Initiative at MIT which aims to use global digital platforms to solve major social problems.